\documentclass[aps,preprint,a4paper]{revtex4-2}
\usepackage{graphicx}
\usepackage{amssymb,amsmath}
\usepackage{nameref}
\usepackage{url}
\usepackage[utf8]{inputenc}
\usepackage[T1]{fontenc}
\usepackage[english]{babel}

\begin{document}

\title{Size Effects in Periodic Metamaterials}
\author{Victor V. Gozhenko}
\affiliation{Institute of Physics, Natl. Acad. of Sciences of Ukraine, 46 Nauky Ave., Kyiv 03680, Ukraine}
\email{vigo@iop.kiev.ua}

\begin{abstract}
The optical properties of periodic electromagnetic metamaterials are considered as functions of their relative unit cell size $d / \lambda$. The reflection $R$ and transmission $T$ coefficients are numerically calculated for some realistic metamaterials in a wide range of their relative unit cell size values that comprises different operating regimes of the metamaterials. Peculiarities in $R$ and $T$ behavior are discussed and the causes of those peculiarities are outlined. The obtained results support the opinion on inapplicability of the very homogenization concept to metamaterials whose unit cell size is comparable to the incident wavelength, in contrast to some previously published results.
\end{abstract}

\maketitle

\section{Introduction}

\begin{figure}[b]
 \begin{center}
  \includegraphics*[width=0.75\textwidth]{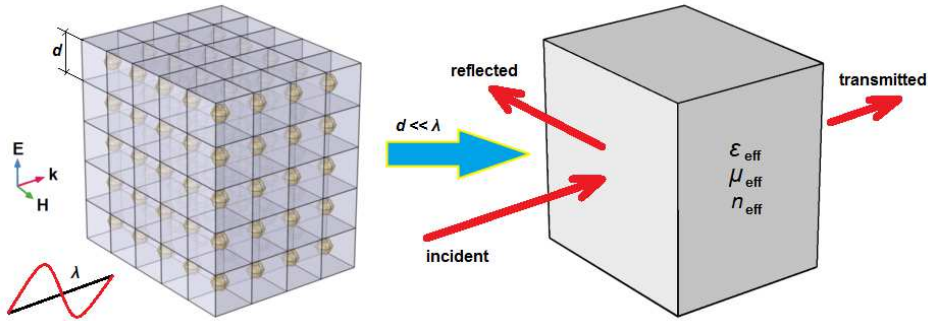}
   \caption{\label{fig1}The concept of metamaterials homogenization. A periodic metamaterial with unit cells of size $d$ is lit by an incident electromagnetic wave whose wavelength is $\lambda$. If $d \ll \lambda$, then the incident wave cannot ``feel'' the metamaterial inhomogenities, and metamaterial behave itself like a continuous and homogeneous medium, whose parameters are $\varepsilon_{\textrm{eff}}$, $\mu_{\textrm{eff}}$, $n_{\textrm{eff}}$. Representation of a metamaterial by the corresponding homogeneous medium is correct if the optical properties (e.g., reflectance and transmittance) of the metamaterial and the medium are the same.}
 \end{center}
\end{figure}

Most of the electromagnetic metamaterials are periodic structures, and their unit cells consist of artificial inclusions designed to get a specific electromagnetic response (e.g., negative refraction or selective reflectivity) of the metamaterial sample as a whole. Periodic metamaterial can be treated as a continuous and homogeneous medium if its unit cell size $d$ (the lattice constant) is much smaller then its operating wavelength $\lambda$, $d \ll \lambda$. In such a case, the metamaterial can be characterized by its effective parameters---the effective permittivity $\varepsilon_{\textrm{eff}}$, permeability $\mu_{\textrm{eff}}$, and index of refraction (the refractive index) $n_{\textrm{eff}}=\sqrt{\varepsilon \mu}$, see Fig.~\ref{fig1}.

Calculation of the effective parameters values for a given metamaterial (with given shape, size, and material of its inclusions, as well as the size and geometry of its unit cell) is important, for example, for predicting the optical properties the metamaterial will reveal in experiments and applications, and is based on calculating the local electric and magnetic fields within the unit cell and proper averaging of those fields over the cell. Sometimes---in case of simple inclusions---it can be done analytically; in general case, numerical computations are required.

To facilitate calculation of the effective parameters of periodic metamaterials, a number of homogenization theories and methods were proposed (see, e.g., Refs.~\onlinecite{Pendry1999, Smith2006, Tsukerman, Gozhenko2013, Yoo2019, Rybin2022}). Those methods differ from each other by, particularly, the way they calculate the average values of the local fields in the metamaterials.

In metamaterials applications, condition $d\ll\lambda$ (say, $d =0.01 \lambda$) is not always met. For example, optical applications (where $\lambda \approx 400\ldots800$ nm) implies that the unit cell size should be of the order of 50 nm or less. However, metamaterials with $d =200\ldots300$ nm (i.e., $d/ \lambda \approx 0.25\ldots0.5$) are often used there because the less the unit cell size, the more expensive manufacturing process of the metamaterial sample. On the other hand, some applications---most notably the negative index of refraction---require for a metamaterial to work in the resonant regime (where the negative $n$ is achieveable) meaning $d/ \lambda \approx 0.5\ldots1.0$. Therefore, some authors tried to elaborate homogenization methods suitable for an extended range of $d/ \lambda$ values, and not only for small $d$. Some of them believe that their methods are valid for metamaterials with substantial or even arbitrary unit cell size (e.g., Refs.~\onlinecite{Pendry1999, Tsukerman, Rybin2022}).

Strictly speaking, any homogenization method can give plausible results for metamaterials working in the long wavelength (quasistatic) regime only, where the condition $d \ll \lambda$ is satisfied. In the opposite case of short (relative to the unit cell size) waves, $d \gg \lambda$, the very homogenization concept should fail, and metamaterials cannot be treated as homogeneous media. In this case, propagation of incident waves through a metamaterial obeys the geometrical optics laws, and reflection of an incident wave from the metamaterial inclusions plays a crucial role. Last, in the intermediate regime, where $d \approx \lambda$, homogenization methods should not work since they do not account for the diffraction effects (e.g., the Bragg's reflection) which are significant in this case.

Earlier \cite{Gozhenko2013}, it was shown that different homogenization methods give more and more diverging results as the relative unit cell size of metamaterials increases from zero to approximately $d/ \lambda = 0.4$. At larger $d/ \lambda$ values, calculations of homogenized effective parameters of a metamaterial can still formally be performed, but those parameters cannot describe correctly the optical properties of the metamaterial.

In the present paper, the optical properties of periodic metamaterials are considered in a wider range of their relative unit cell sizes $d/ \lambda$, and the effects of the cell size on the optical behaviour of the matematerials is discussed in more details.

\section{Basic formulae\label{sec:formulae}}

We are interested in calculating the observable quantities---transmittance $T$ and reflectance $R$ of a metamaterial---which are dimensionless coefficients defined as
$$
T = \frac{I_{\textrm{tr}}}{I_0},\qquad
R=\frac{I_{\textrm{ref}}}{I_0},
$$
where $I_0, I_{\textrm{tr}}, I_{\textrm{ref}}$ are the intensities of incident, transmitted through the metamaterial, and reflected waves. The intensities are the energy flux densities of the corresponding waves and can be calculated as time-averaged values of the Poynting vector $\mathbf{S}$ of those waves (indices are omitted below for simplicity):
$$
I = \langle\mathbf{S}\rangle = \frac{1}{\tau} \int\limits_t^{t+\tau}\mathbf{S}(t') dt'
$$
In case of monochromatic incident plane wave
\begin{gather}
\mathbf{E}=\mathbf{E}_0 e^{[i(\mathbf{k}\cdot \mathbf{r} - \omega t)]},\\
\mathbf{H}=\mathbf{H}_0 e^{[i(\mathbf{k}\cdot \mathbf{r} - \omega t)]}
\end{gather}
all the waves involved are also monochromatic and their intensities can be calculated from
$$
I=\frac{1}{2} \textrm{Re}(\mathbf{E}\times\mathbf{H^*}),
$$
where the asterisk denotes the complex conjugation.
From the energy conservation law applied to the interaction of electromagnetic waves with a lossy medium, it follows
$$
T+R+A=1,
$$
where $A$ is the absorptance (the absorption coefficient) of the medium which define the rate of electromagnetic energy absorption inside it. If $T$ and $R$ values for a medium are known (i.e., are experimentally measured or theoretically calculated), its absorptance can be found as
$$ A = 1 - T - R. $$

\section{Numerical results and discussion\label{sec:results}}

Numerical simulations are carried out for metamaterials with cubic lattice and consisted of inclusions of various shapes that are often used in metamaterial science and applications---spheres, rods, Split-Ring Resonators (SRRs), and $\Omega$-like inclusions (``omegas''). Optical reflectance $R$ and transmittance $T$ of the simulated metamaterials are calculated in a wide range of their relative unit cell size $d/ \lambda$ at several values of the incidence angle $\theta$. Calculations of $R$ and transmittance $T$ are complemented with the local $E$ field distributions across the unit cells in different regimes the metamaterials operate in. All the calculations are performed with COMSOL Multiphysics software. Presented below are exemplary calculation results.

\begin{figure}
 \begin{center}
  \includegraphics*[width=0.75\textwidth]{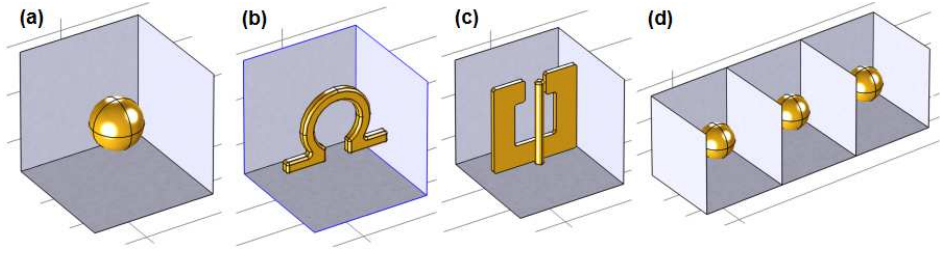}
   \caption{\label{fig2}Unit cells of the simulated metamaterials: (a) a monolayer of spherical particles; (b) a monolayer of omegas; (c) a monolayer of SRRs and rods; (d) a triple layer of spherical particles. All the inclusions are made of gold, and the size of the individual unit cells is $d=500$ nm.}
 \end{center}
\end{figure}

Schematics of the unit cells of the simulated metamaterials are shown in Fig.~\ref{fig2}. The materials are primarily infinite monolayers of thickness $d$ and that in Fig.~\ref{fig2}d is a triple layer of thickness $3d$. The unit cell size $d=500$ nm is the same for all the materials and remains unchangeable in all the calculations. Variations in the relative unit cell size $d/ \lambda$ are made by changing the wavelength $\lambda$ of the incident wave.

\begin{figure}
 \begin{center}
  \includegraphics*[width=0.75\textwidth]{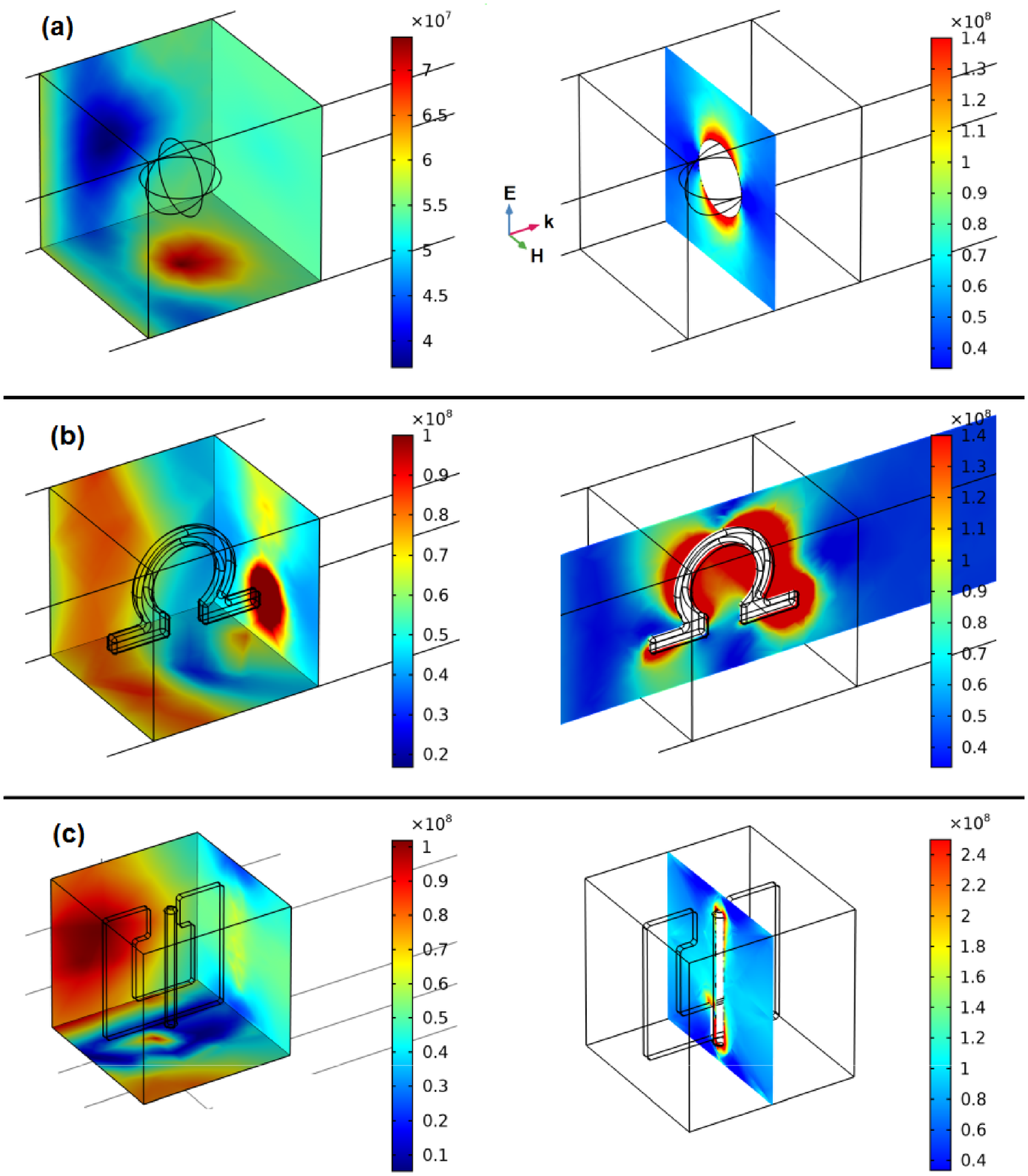}
   \caption{\label{fig3}Local distributions of the absolute value of the electric field $\mathbf{E}$ inside the unit cells of the metamaterials shown in Fig.~\ref{fig2}a--c. Left column, the field on the unit cell boundaries; right column, the field in the middle plane of the cells. $\lambda = 1000$ nm, normal incidence ($\theta = 0$). Directions of $\mathbf{E}$ and $\mathbf{H}$ vectors of the incident wave are depicted in panel (a). Distribution of $|\textbf{E}|$ allows one to easily determine those places where the electric energy is concentrated.}
 \end{center}
\end{figure}

From Fig.~\ref{fig3} one can see that all the inclusions give a prominent response to the incident wave even at normal incidence, the response of spherical inclusions being dipole-like (see panel (a), the field distribution in the middle plane), while those of ``SRRs with rods'' system and ``omegas'' are more tricky.

\begin{figure}
 \begin{center}
  \includegraphics*[width=0.75\textwidth]{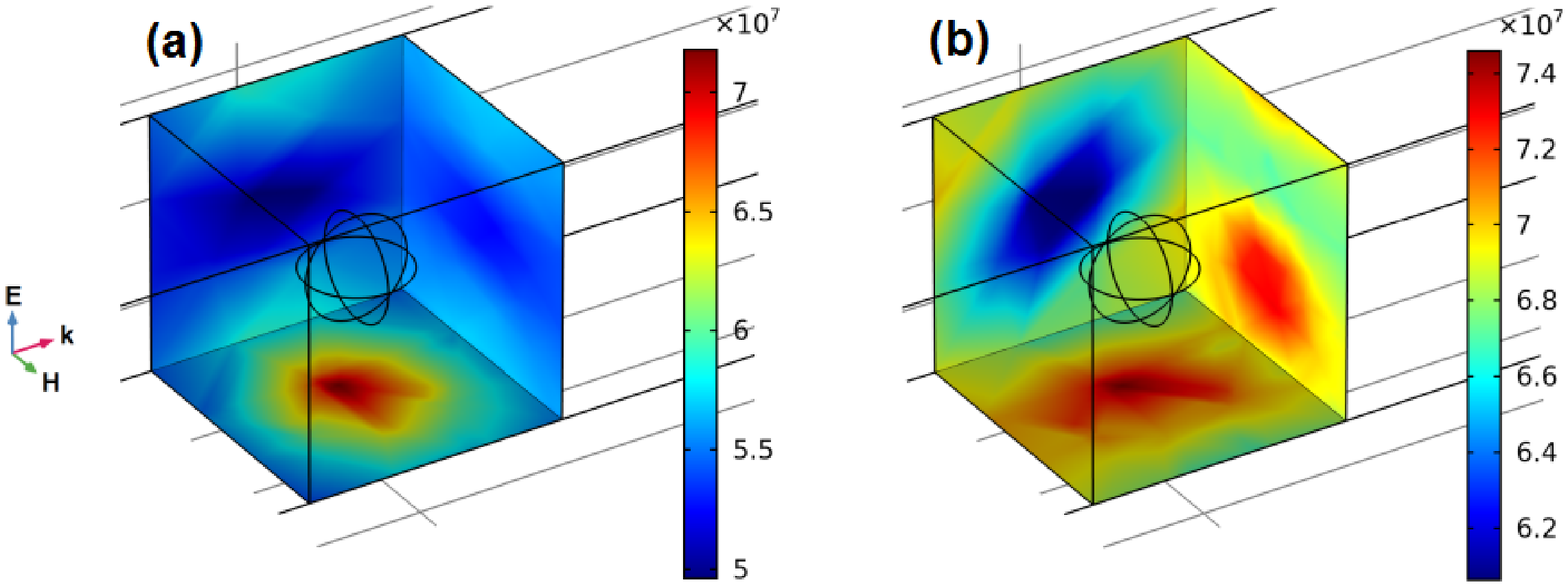}
   \caption{\label{fig4}Local distribution of $|\mathbf{E}|$ over the unit cell boundaries in a monolayer of golden spheres at $\lambda = 5000$ nm. (a) $\theta = 0$; (b) $\theta = 45^{\circ}$.}
 \end{center}
\end{figure}

The incidence angle $\theta$ also affects the electromagnetic field distribution inside metamaterials: even for simple spherical inclusions in the quasistatic regime ($\lambda = 10d$) the electric fields in the unit cell differ substantially at normal and oblique incidence, see Fig.~\ref{fig4}.

\begin{figure}
 \begin{center}
  \includegraphics*[width=0.75\textwidth]{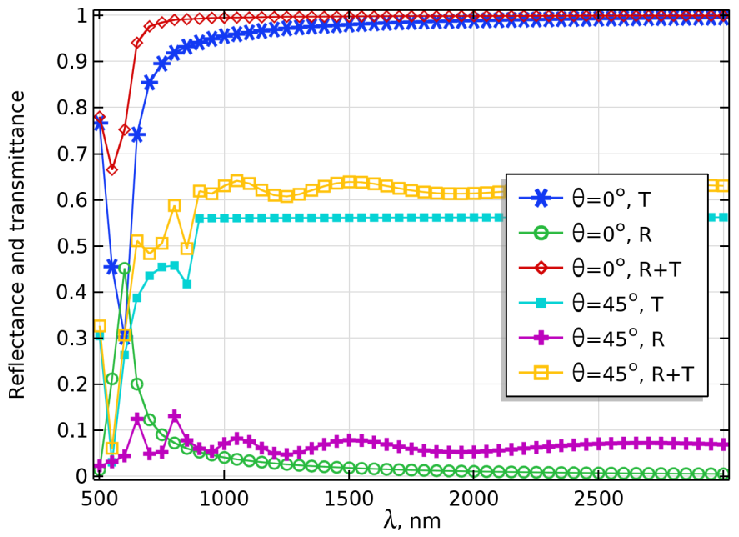}
   \caption{\label{fig5}Transmission and reflection spectra of a monolayer of golden spheres at normal and oblique incidence, $\theta = 0$  and $\theta = 45^{\circ}$.}
 \end{center}
\end{figure}

\begin{figure}
 \begin{center}
  \includegraphics*[width=0.75\textwidth]{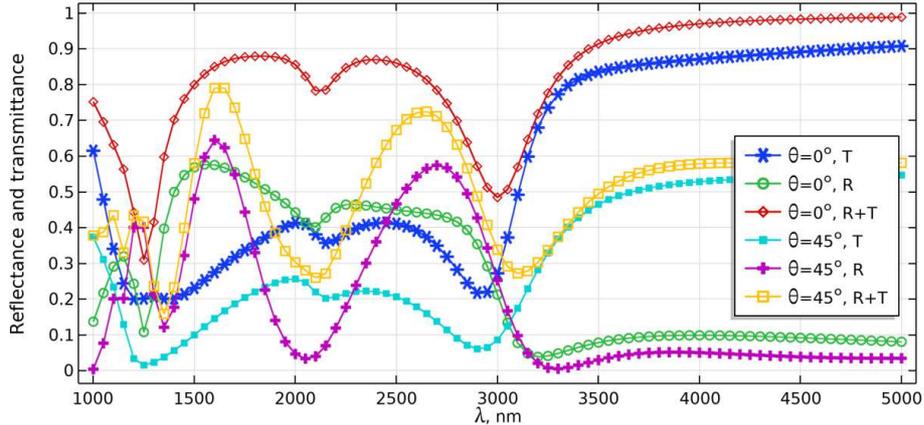}
   \caption{\label{fig6}Transmission and reflection spectra of a monolayer of golden inclusions ``SRRs plus rods'' at normal and oblique incidence, $\theta = 0$  and $\theta = 45^{\circ}$.}
 \end{center}
\end{figure}

\begin{figure}
 \begin{center}
  \includegraphics*[width=0.75\textwidth]{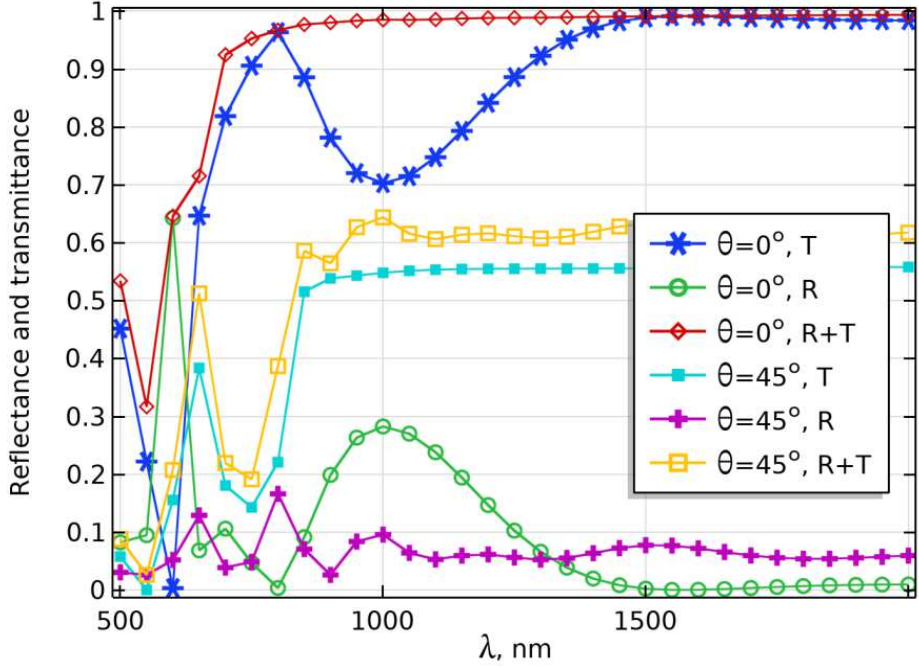}
   \caption{\label{fig7}Transmission and reflection spectra of a triple layer of golden spheres at normal and oblique incidence, $\theta = 0$  and $\theta = 45^{\circ}$.}
 \end{center}
\end{figure}

Shown in Figs.~\ref{fig5}--\ref{fig7} are the transmission and reflection spectra numerically calculated for the metamaterials depicted in Fig.~\ref{fig2}a,c,d.

According to Fig.~\ref{fig5}, at normal incidence of a long enough wave with $\lambda = 3000$ nm (which is six times the unit cell size $d$) onto the monolayer of golden spheres, the metamaterial behaves itself as a transparent sheet: its reflection coefficient $R$  in this regime is close to zero, and its transmission coefficent $T$ is near unity. If $\lambda$ decreases and approaches the unit cell size $d$, the metamaterial gradually looses its transparency and becomes more and more reflective. The reflection coefficient has its maximum $R = 0.45$ at $\lambda = 1.2 d = 600$ nm, and the transmission coefficient is minimal ($T = 0.3$) at this point. Further decrease of $\lambda$ results in monotonic increase of the material transparency (with the peak value $T \approx 0.77$ at $\lambda = d = 500$ nm) and monotonic decrease of its reflectance up to $R \approx 0$. The radical change in the behavior of $R$ and $T$ at $\lambda \le 600$ nm (when $d/ \lambda \ge 0.83)$ is probably due to the onset of the diffraction and interference effects in the periodic metamaterial.

Note that the sum $T+R$ becomes distinctly less then unity at $\lambda \le 650$ nm, or $d/ \lambda \ge 0.77$. It means that light absorption by the metamaterial is substantial in this case. The minimum value of the absorption coefficient $A = 1 - T + R \approx 0.67$ is observed at $ \lambda = 550$ nm, which implies that about one third of the incident energy flux is absorbed by the metamaterial inclusions. The absorption can be ascribed to electric currents induced in individual inclusions (golden spheres) by the incident wave as well scattered waves from neighboring particles.

Analogous behavior is observed at oblique incidence (see the respective curves in Fig.~\ref{fig5} for the case of $\theta = 45^{\circ}$): with decreasing $\lambda$, the region of insufficient changes in $R$, $T$ passes (starting from $\lambda \approx 1.8 d = 900$ nm) to the region of their abrupt changes and oscillating behavior. Note, however, that the metamaterial in this case is semi-transparent even at large $\lambda$: $R+T \approx 0.62$ at $\lambda = 3000$ nm.

The plots in Fig.~\ref{fig6} refer to the monolayer of golden inclusions ``SRR plus rod'' and have three distinct regions at both normal and oblique incidence. With decrease in $\lambda$ from its maximum value 5000 nm to approximately 3200 nm, the optical properties of the monolayer change monotonically. Further, up to $\lambda \approx 1400$ nm, there is the region of oscillations, where one can observe peaks and dips of $R$ and $T$ at wavelenghts that are nearly multiples of $d=500$ nm. Those peaks and dips can be ascribed to the grating resonances occurred in periodic systems as a result of interaction between their structural elements excited by the incident wave. With further decrease in $\lambda$, the oscillations of $R$ and $T$ become more chaotic. As in the case of spherical inclusions, the layer of SRRs and rods looks translucent even at large enough wavelenghts: $T+R<0.6$ at $\lambda = 10 d= 5000$ nm.

The peculiarities in the optical behavior of monolayers of golden spheres and SRRs with rods can also be observed in a thicker triple layer of golden spheres, see Fig.~\ref{fig7}. In this case, notice the two peaks in $R$ (at normal and oblique incidence) and the dip in $T$ (at normal incidence) which are all located exactly at $\lambda = 1000$ nm, which is two times the lattice constant $d$. Obviously, they can be ascribed to the above mentioned grating resonances.

\section{Conclusions\label{sec:conclusions}}

The obtained results confirm the opinion \cite{Gozhenko2013} that any metamaterial homogenization method should be used with care in the intermediate operating regimes, when the metamaterial unit cell size is of the order of the operating wavelength. The value of the relative unit cell size $d/ \lambda$ at which homogenization methods fail to predict the optical properties of periodic metamaterials depends on the geometry and material parameters of their inclusions. For the metamaterials we considered here, abrupt and substanial changes in the optical properties (as compared to their longwavelength values) occur at different values of $d/ \lambda$: near 0.56 for the monolayer of golden spheres at oblique incidence, 0.4 for the triple layer of golden spheres at normal incidence, and near 0.16 for the monolayer of SRRs with rods. For larger $d/ \lambda$ values, a crucial role in the optical properties formation play the diffraction and interference effects in the metamaterials, so the properties exhibit an oscillating behavior which cannot be predicted within the homogenization concept.


\begin{thebibliography}{99}

\bibitem {Pendry1999}Pendry J.B., Holden A.J., Robbins D.J. and Stewart W.J. IEEE Trans. Microw. Theory Tech. {\bf 47} 2075--84 (1999).
\bibitem {Smith2006}D. Smith and J. Pendry, J. Opt. Soc. Am. B  {\bf 23} 391-403 (2006).
\bibitem {Tsukerman}I. Tsukerman, J. Opt. Soc. Am. B {\bf 28} 577--86 (2011).
\bibitem {Gozhenko2013}V.V. Gozhenko, A.K. Amert, and K.W. Whites , New J. Phys. {\bf 15} 043030 (2013).
\bibitem {Yoo2019}S. Yoo et al., Nanophotonics 8 (6) 1063--1069 (2019).
\bibitem {Rybin2022}O. Rybin and V. Khardikov, Optik {\bf 268} 169768 (2022).

\end{thebibliography}
\end{document}